# Empirical Study of Traffic Velocity Distribution and its Effect on VANETs Connectivity

Sherif M. Abuelenin
Department of Electrical Engineering
Faculty of Engineering, Port-Said University
Port-Fouad, Port-Said, Egypt
s.abuelenin@eng.psu.edu.eg

Adel Y. Abul-Magd
Department of Mathematics
Faculty of Science, Zagazig University
Zagazig, Egypt
aamagd@gmail.com

*Abstract*— **In this article we use real traffic data to confirm that vehicle velocities follow Gaussian distribution in steady state traffic regimes (free-flow, and congestion). We also show that in the transition between free-flow and congestion, the velocity distribution is better modeled by generalized extreme value distribution (GEV). We study the effect of the different models on estimating the probability distribution of connectivity duration between vehicles in vehicular ad-hoc networks.**

## I. Introduction

In vehicular ad-hoc networks (VANETs), the connection duration between any two vehicles depends on their relative velocity. It is widely accepted that velocity distribution of vehicles in highways follows normal distribution [1]. Krbalek [2] has theorized that it was possible to model steady-states of vehicle traffic in terms of the thermal-equilibrium properties of a Dyson's gas exposed to a heat reservoir, where the particles interact by a short-ranged power-law. Vehicle velocities in this model satisfy a Gaussian (normal) distribution.

According to Kerner's traffic theory [3], the stationary state of highway traffic can be either free-flowing or congested. Traffic congestion is divided into two phases; synchronized flow and wide moving jams. The transition between free-flow and congestion is characterized by the co-existence of the two phases [4]-[7]. As Fig. 1. shows, the transition from free-flow to traffic jam starts with the introduction of small (clustered) jams in the road, with vehicles travelling freely in between such jams. Afterwards, the jams widens up until they merge into one large traffic jam.

In this article we use empirical vehicle data, taken from measurement in the Berkeley Highway Laboratory (BHL) project [8], to confirm that vehicle velocities follow Gaussian distribution in steady state traffic regimes (free-flow, and wide moving jam). We also show that in the transition between free-flow and congestion, velocity distribution is better modeled by the generalized extreme value distribution (GEV).

The rest of the article is organized as follows; in section II we discuss the distribution of vehicles velocities. In section III we study how the different distributions affect connection

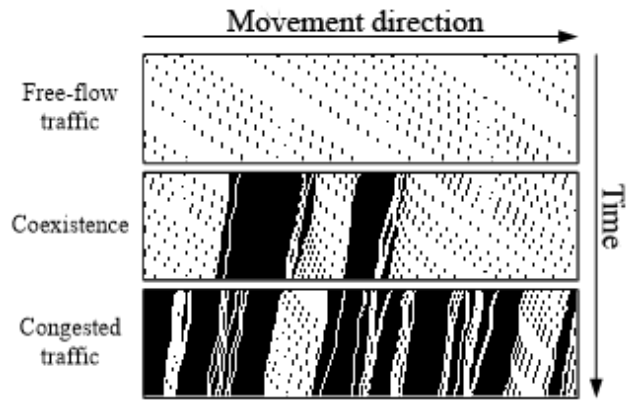

Fig.1. Cellular Automata model of traffic phases, [4]

duration in VANETs, and we provide our results. The paper is concluded in section IV.

## II. Velocity Distribution

### A. Traffic Data

The vehicles data used in this paper are obtained from Berkeley Highway Laboratory project. The data are collected using dual loop sensor stations installed on the five lane interstate I-80 road in California. Each station is a pair of inductive loop detectors, one upstream and one downstream in the same lane. The two sensors are six feet across, and their centers are twenty feet apart. The project provides individual raw vehicle data. Each vehicle stream data file record indicates a matched pair of upstream and downstream transitions from a specific lane at a specific station. The arrival instants are accurate within 1/60 of a second. This enables one to calculate each vehicle speed and the spacing between successive vehicles passing in the same lane [9].

We compute the velocity of each vehicle in the road for a period of one day (24 hours). Fig. 2 shows the mean velocity in every hour of the 24-hour period. Different time periods are selected to represent different traffic regimes, with each time period lasting for one hour. We empirically find the cumulative density functions (CDF) of each, and fit it to a Gaussian CDF, to confirm Krbalek's theory. Fig. 3 shows the empirical and Gaussian resulting CDFs. The figure shows that the velocity distribution closely resembles Gaussian distribution, except in the transition between free-flow and congestion (3:00pm, as well as 5:00pm, not shown). Empirical CDF for the other steady state periods that was not shown here also resembles Gaussian distribution.

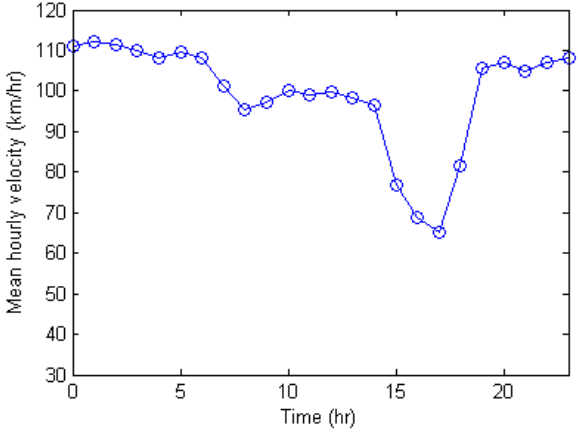

Fig.2. Mean hourly velocity in 24 hour period

*B. Gaussian Distribution*

Krbalek et al [2], [10], [11] use one-dimensional thermodynamical particle gas to predict microscopic structure in traffic flows and consequently compare to the relevant traffic data distributions. Justification for the approach is provided in [2] and [12] where it is proved that equilibrium solution of certain family of the particle gases (exposed to a heat reservoir) is a good approximation for steady-state solution of driven many-particle system. The thermodynamic approach leads to the assertion that velocity $\varpi$ of particles is Gaussian distributed according to the following probability density function

$$f(v) = \frac{1}{\sqrt{(2\pi\sigma^2)}} e^{-\frac{(v-\mu)^2}{2\sigma^2}} \qquad (1)$$

*C. Alternative Distributions*

The generalized extreme value distribution was shown [13] to model vehicles headway distribution in transition phase. Also the lognormal distribution successfully modeled traffic headway distributions in the same phase [14]. We applied both distributions to our data, and we found that they also provide better fits for velocity distribution in the transition regime (see Fig 3.b). Both distributions were fitted to the empirical data with root mean square error (RMSE) values that are considerably lower that of the corresponding normal distribution. For both steady state traffic regimes (free-flow, and jam), neither the GEV nor the lognormal distribution provided a good fit compared to Gaussian distribution. Since the focus of this article is to study the impact of velocity distribution on VANETs connectivity (as discussed in the next section), which requires studying the distribution of relative velocity between vehicles. We limit our discussion in this article to GEV distribution, and will keep the study of lognormal distribution to future work. Studying the distribution of the difference between two lognormal random variables is a rather complex process [15].

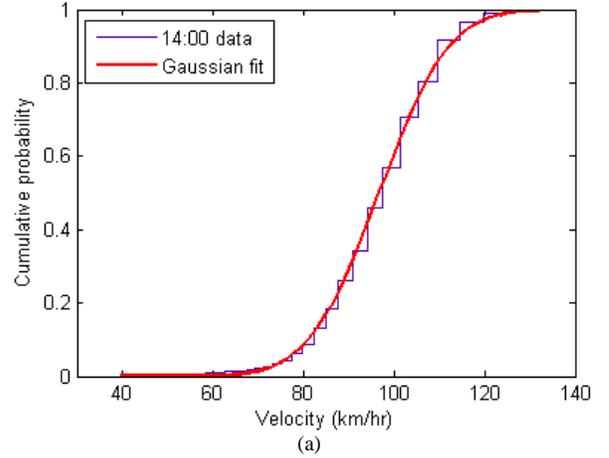

(a)

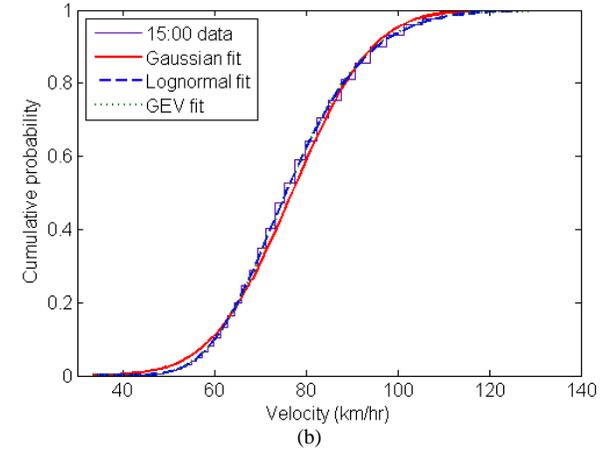

(b)

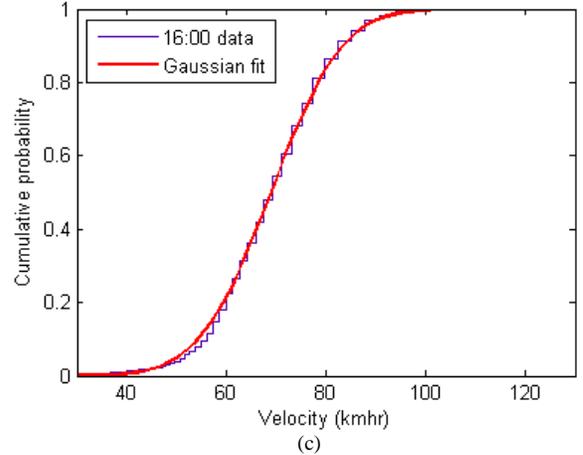

(c)

Fig.3. Empirical and Gaussian CDF plots of traffic velocities; (a) 2:00 pm, (b) 3:00 pm, (c) 4:00 pm.

GEV distribution is a family of probability distributions that combines the Gumbel, Fréchet and Weibull distributions. It is considered to be the only possible limit distribution of properly normalized maxima of a sequence of independent and identically distributed (i.i.d.) random variables. The corresponding probability density function is given by

$$f(v) = \frac{1}{\sigma} t(v)^{k+1} e^{-t(v)}, \qquad (2)$$

where

$$t(x) = \begin{cases} \left(1 + (\frac{x-\mu}{\sigma})k\right)^{-1/k} & ; k \neq 0 \\ e^{-(x-\mu)/\sigma} & ; k = 0 \end{cases}$$

for $x > \mu - \sigma/k$ in the case $k > 0$, and for $x < \mu - \sigma/k$ in the case $k < 0$. The density is zero outside of the relevant range.

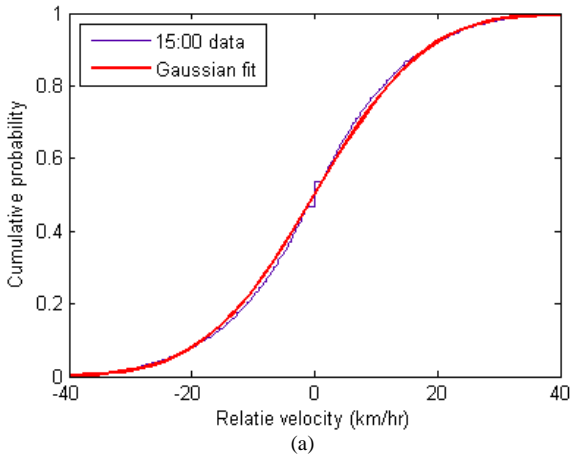

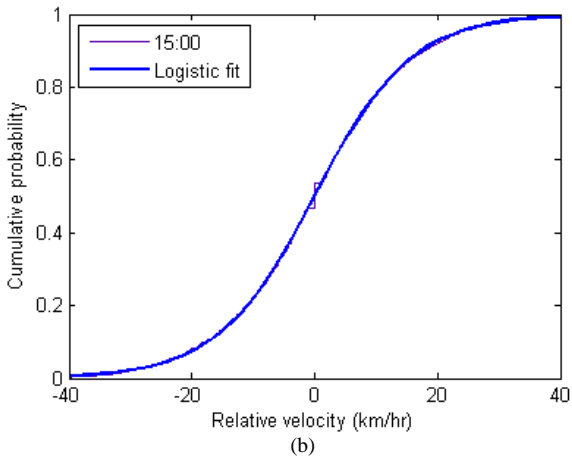

Fig.4. Plots of relative vehicles velocities distribution; (a) Empirical and Gaussian, (b) Empirical and Logistic.

TABLE I. Summary of 3:00 pm velocity distribution fitting

| Velocity, 3:00 pm | GEV | Gaussian |
|---|---|---|
| Model parameters | μ = 71.1 ± 0.1 km/hr<br>k = −0.125 ± 0.003<br>σ = 12.5 ± 0.1 km/hr | μ = 76.9 ± 0.1 km/hr<br>σ = 13.8 ± 0.1 km/hr |
| RMSE | 0.016 | 0.022 |

Table I summarizes the results of fitting the 3:00 pm velocity data to both Gaussian and GEV distributions. The root mean square error between empirical and fitted curve is much less in the case of GEV distribution.

GEV distribution has three parameters; μ for location, σ for scale, and $k$ for shape. The fitting results show that $k$ is close to zero, which mean that the distribution is almost a Gumbel (GEV-type-I) distribution.

We have also compared the empirical velocity distribution in the transitional region with the lognormal distribution. The best-fit results were practically identical with the corresponding fits to the GEV distribution.

In the next section we show the impact of this on connectivity analysis of VANETs.

### III. VEHICULAR AD-HOC NETWORKS

VANETs are special types of mobile ad-hoc networks. In VANETs, vehicles are equipped with transmitters and receivers to enable message dissemination, and information exchange between them. VANETs have highly dynamic topology because of the high mobility of vehicles.

#### A. Relative velocity

In VANETs, connectivity duration between two vehicles is a function of the communication range and the relative velocity between them. Estimating the connectivity duration depends on the distribution of relative velocity. Connectivity duration using the Gaussian distribution model of velocities has been studied (in [1]).

In this section we use our findings to study connectivity duration using GEV distribution.

We have shown earlier that velocity distribution follows a Gumbel distribution in the transition between free-flow and congested traffic. From statistics, it is known that if two random variables follow a Gumbel distribution [16], then the distribution of their difference follows a logistic distribution. The logistic distribution has the following probability density function;

$$f(v) = \frac{e^{-\frac{v-\mu}{\sigma}}}{\sigma\left(1 + e^{-\frac{v-\mu}{\sigma}}\right)^2} = \frac{1}{4\sigma}\operatorname{sech}^2\left(\frac{v-\mu}{2\sigma}\right) \qquad (3)$$

and cumulative distribution function;

$$F(v) = \frac{1}{1+e^{-\frac{v-\mu}{\sigma}}} = \frac{1}{2} + \frac{1}{2}\tanh\left(\frac{v-\mu}{2\sigma}\right) \quad (4)$$

Fig. 4. shows the result of fitting the empirical velocity difference between consecutive vehicles to both Gaussian and Logistic distributions. The figures (as well as the fitting data, summarized in table II) show that indeed the relative velocity distribution is better modeled using Logistic distribution.

We use these results to estimate the probability distribution of the connectivity duration.

*B. Connectivity Duration*

The connectivity duration between two vehicles in a VANET is a function of their relative velocity (the absolute difference between their velocities).

$$t_c = 2R/|\Delta v| \quad (5)$$

where $t_c$ is the connectivity duration, $R$ is the communication range, and $\Delta v$ is the relative velocity between two vehicles. This equation has two solutions; $\Delta v = \pm 2R/t_c$

The random variable $t_c$ is a function of the random variable $\Delta v$. The distribution of $t_c$ can be easily found using the distribution of $\Delta v$ [17]. If the distribution probability density function (PDF) of the velocity difference is known to be $f_v(v)$, (and assuming that the lane separation between vehicles is much less than the headway) then, the distribution PDF of connectivity duration '$t_c$' can be found to be [1];

$$f_{T_C}(t_c) = \frac{2R}{t_c^2}\left(f_v(\frac{2R}{t_c}) + f_v(-\frac{2R}{t_c})\right)u(t_c) \quad (6)$$

where, $u(x)$ is the Heaviside unit step function.

In case the PDF is an even function, (6) is reduced to;

$$f_{T_C}(t_c) = \frac{4R}{t_c^2}\left(f_v(\frac{2R}{t_c})\right)u(t_c) \quad (7)$$

Both suggested velocity difference PDF functions (Gaussian and logistic distributions) are even functions, therefore, (7) holds for our work.

Fig. 5. shows the PDF for connectivity duration for both distributions (Blue for Gaussian, and red for Logistic), R=100m. From the figure, it is shown that, for the same transmission range, the probability of having two vehicles communicating for certain duration is higher when estimated using Logistic distribution. E.g. for transmission range of 100m, using the logistic distribution estimates that approximately 52.8% of the vehicles will be connected for a duration of more than 80 seconds, while using the Gaussian distribution estimates that only 47.8% of the vehicles will be connected for more than 80 seconds.

TABLE II. Summary of relative velocity fittings

| Relative velocity, 3:00 pm | Logistic | Gaussian |
|---|---|---|
| Mean | 0.00 ± 0.01 | 0.00 ± 0.02 |
| Standard deviation | 7.95 ± 0.01 | 13.42 ± 0.04 |
| RMSE | 0.004 | 0.009 |

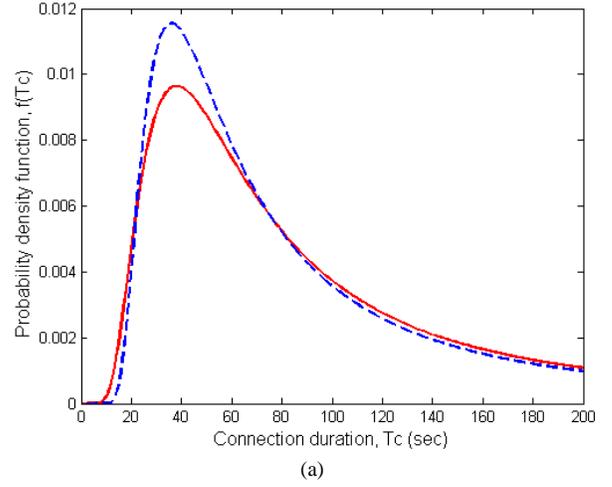

(a)

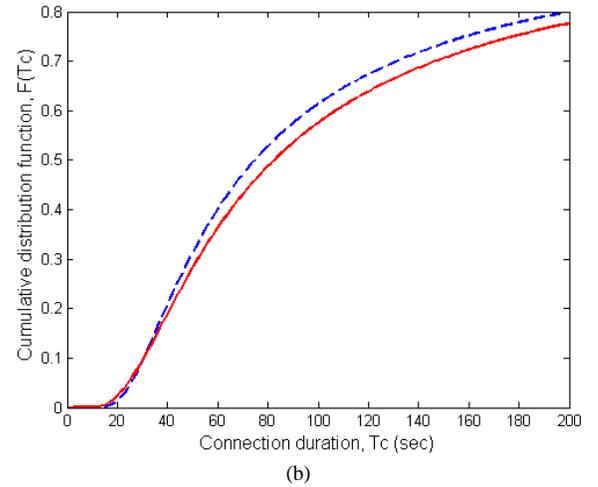

(b)

Fig.5. PDF (a) and CDF (b) of Connection duration using Gaussian distribution (blue, dashed) and Logistic distribution (red, solid).

## IV. CONCLUSIONS

In this article we used real traffic data to confirm that vehicle velocities follow Gaussian distribution in steady state traffic regimes (free-flow, and congestion). We also showed that in the transition between free-flow and congestion velocity distribution is better modeled by Generalized extreme value distribution. We showed that this results in different estimation of connection duration between two vehicles in a VANET.

We limited the work presented here to vehicles travelling down the same direction. In future work we will consider vehicles travelling in opposite directions of the highway.